\UseRawInputEncoding
\documentclass[dvips,twocolumn,aps,amssymb,10pt, showkeys]{revtex4}

\expandafter\let\csname equation*\endcsname\relax
\expandafter\let\csname endequation*\endcsname\relax
\everymath=\expandafter{\the\everymath\displaystyle}

\usepackage{amsmath}
\usepackage{amssymb}
\usepackage{color}
\usepackage{verbatim}
\usepackage{array}
\usepackage{graphicx}
\usepackage{epsfig}
\usepackage{bm}
\usepackage[ragged]{sidecap}
\usepackage{dsfont}
\usepackage{ulem}
\usepackage{mathrsfs}
\usepackage{setspace}

\usepackage{subfigure}
\usepackage{placeins}

\newcommand{\sss}{\smallskip}

\newcommand{\ket}[1]{\mathinner{|{#1}\rangle}}
\newcommand{\braket}[2]{\langle #1|#2\rangle}

\newcommand{\dd}{\mathrm{d}}

\begin{document}
\preprint{\hfill\parbox[b]{0.3\hsize}{ }}

\title{
Quantum Measurements, faster-than-light communication, minimum size for the detector implied by quantum mechanics
}

\author{
F. Fratini${}^1$\footnote{
	\begin{tabular}[t]{ll}
	E-mail addresses: & fratini.filippo@gmail.com\\
	&fhb191379@fh-vie.ac.at
	\end{tabular}
	}
	\vspace{0.3cm}
}

\affiliation{\it ${}^1$ 
University of Applied Sciences BFI Vienna, Wohlmutstra\ss e 22, A-1020 Wien, Austria
}

\date{\today}

\begin{abstract}
We describe an interference setup which seems to defy the No-Communication theorem. We discuss its potential flaw and derive a relation implied by quantum mechanics that limits the size of the detector given the particle's state width.
\end{abstract}


\maketitle

%

\section{Introduction}
\label{sec:intr}
The debate about the fuzziness and the meaning of quantum mechanics traces back to Albert Einstein and Niels Bohr, in the early days of the twentieth century. In a letter to Max Born from 1926, Einstein famously lamented \cite{Pais1982} `{\it Quantum mechanics is very impressive. But an inner voice tells me that it is not yet the real thing. The theory produces a good deal but hardly brings us closer to the secret of the Old One. I am at all events convinced that He does not play dice.}' Bohr, on the other hand, did not share such a discomfort. Bohr was one of the strong promoters of the quantum philosophy, one of the core principles of which is that any knowledge about an atomic system one may obtain will always involve a peculiar indeterminacy \cite{Bohr1949}. 
At the 5th Solvay conference in 1927, with Einstein's unease, Bohr and his followers declared that the quantum revolution was to be considered concluded \cite{Concluded}. Many years of quantum success followed. Year after year, by blindly applying quantum rules, physicists have been led to the correct answer of physical problems. 

\sss

The quantum success culminated in the second half of the twentieth century. In 1964 John Bell showed that there is a measurable difference between any possible local theory of nature and quantum mechanics \cite{Bell_theorem}. During the following years, Bell's theorem was experimentally tested, thus confirming the correctness of quantum mechanics \cite{Aspect1982}. Since then, `{\it shut up and calculate!}' has been the common reaction to those few who happened to be still questioning about the fuzziness of quantum mechanics \cite{Mermin1989}.

\sss

Oddly enough, three people that perhaps mostly contributed to quantum mechanics --- namely Albert Einstein, who strongly helped the quantum birth via his works on light quanta; Erwin Schr\"odinger, who donated the hearth to quantum mechanics via the equation that bears his name; and John Bell, who provided the theorem whose experimental realization gave quantum mechanics its glorification --- were utterly dissatisfied with it. Schr\"odinger commented on quantum mechanics with \cite{Schroedinger} `{\it I don't like it, and I am sorry I ever had anything to do with it.}' Similarly, Bell expressed his discomfort with quantum mechanics several times quite explicitly \cite{Bell_incomplete}: ` [...] {\it quantum mechanics is, at the best, incomplete.}' 

\sss

Notwithstanding the success of quantum mechanics, the debate about its meaning and counterintuitivity has never been discontinued. More recently, Steven Weinberg gave a brilliant account of the current dissatisfaction \cite{Weinberg2017}, while more and more dissenting voices are being heard \cite{Smolin2019, Hossenfelder2018, Woit2007}.
At the 25th Solvay conference on Physics in October 2011, when discussing the longstanding problems of quantum mecahnics, Alain Aspect put forward a call \cite{Solvay2011}: `{\it I am looking forward with a great interest to some kind of theorem providing the possibility for a test, because at the end of the day Nature is our judge.}'. All of those testify that there is still widespread interest in better understanding the longstanding conundrums of quantum mechanics.

Along the lines of the EPR work \cite{EPR}, several thought experiments have been proposed throughout the last century, aimed at recording the quantum spooky action at distance, as Einstein named it \cite{Spooky1971}, so to allow faster-than-light communication \cite{GhirBook, error, Herbert1982}. Some of them have been found to be profoundly useful for the development of quantum science \cite{Peres2003}. The aim of those attempts is to defy the No-Communication (NC) theorem \cite{Ghirardi1980}, which forbids instantaneous communication between distant observers. Why would someone try to defy a theorem? Mainly because the theory of quantum measurement is not considered a valid theory by some (Bell was one of them).
Furthermore, exact quantum evolution is not possible for most of case studies. Approximations and ad-hoc rules are widely used when trying to describe phyical problems and experimental settings. All those could be thought to be helpful in defying the NC theorem.

This paper is another attempt. We use two hypotheses (to be read in Sec. \ref{hyp}) which seem reasonable in light of recent experiments. By using those we conceive an experimental setup which allows for faster-than-light communication. We do not claim to say that our setup really provides faster-than-light communication; rather we aim to stimulate thinking and comparison between what the theory says and what experiments find. We discuss an intuitive explanation which would validate the conclusions. On the other hand, we discuss a more rigorous description in terms of Quantum Mechanics which would invalidate the conclusions. However, we highlight that this latter seems not to represent experiments. Hence the question remains open.

\begin{figure*}[t]
\includegraphics[scale=0.7]{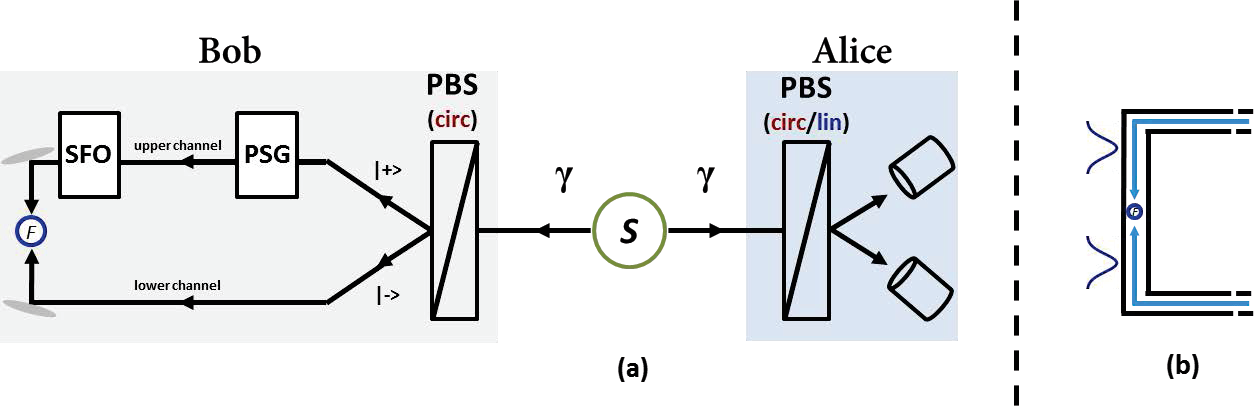}
\caption{(a) The experimental setup of the gedanken experiment. A source of maximally entangled photon pairs ($S$) sends entangled photons to Alice and Bob. Alice uses a polarization beam splitter (PBS), which can be aligned in the circular or linear basis at Alice's will, with the aim to send information to Bob. On the other hand, Bob, with the aim to understand Alice's message, uses an interferometer setup, which comprises a PBS in the circular basis, a phase shift generator (PSG), a spin flip operator (SFO), a set of mirrors to let the beams interfere head-on at the point $F$, where a small photodetector is placed.
(b) The interference region.}
\label{fig:GE}
\end{figure*}

\section{Hypotheses}
\label{hyp}
Our assumptions are:
\begin{enumerate}
\item photon detectors can be made as small as needed,
\item photon detectors are ideal: when a photon crosses the detector entirely, the detector clicks with 100\% probability.
\end{enumerate}
These hypotheses look realistic, if one uses photon transport via nanometric waveguides, where an artificial atom can be therein embedded \cite{JPhilippe2014}, and be used as detector. Atoms are typically sensible to wavelengths that are orders of magnitude larger than the atom size; atoms can be therefore considered point-like, as compared to photon's wavelength to which they are sensible.
Furthermore 100\% efficient photo-detection is not only a theoretical abstraction, but is also nowadays experimentally achieved in nanometric waveguides thanks to the enhanced atom-photon coupling that the waveguide provides \cite{Marsili2013, Nguyen2018, Chang2018, Chonan2014}. As a consequence of this, for example, the reflection coefficient approaches unity, for photons which are in resonance with the embedded atom \cite{Fratini2014, Valente2012}. Upon reflection, the photon acquires a $\pi$ phase shift, thereby leaving a trace for experiments to record \cite{Fratini2014}, thus testifying that the photon has been `seen' by the atom.

\section{The thought experiment}

\subsection{The set up}
\label{sec:SetupGE}

The thought experiment we consider is showed in Fig. \ref{fig:GE}(a). We suppose to have a source ($S$) which produces maximally entangled photon pairs at a fixed rate, such as 100 photon pairs per second. We chose photons for simplicity, but any other quantum system with two spin states would similarly work. We suppose the quantum state of each photon pair be defined as
\begin{equation}
\begin{array}{lcl}
\ket{\gamma\gamma}&=&\frac{1}{\sqrt{2}}\big( \ket{+-} - \ket{-+} \big)\\[0.4cm]
&=&\frac{i}{\sqrt{2}}\big( \ket{yx} -\ket{xy} \big)~,
\end{array}
\label{eq:State}
\end{equation}
where $\ket{+}$, $\ket{-}$ are circularly polarized states, while $\ket{x}$, $\ket{y}$ are linearly polarized states. In the last step we used the relations that link circular and linear polarizations \cite{Rose}:
\begin{equation}
\begin{array}{lcl}
\ket{x}&=&\frac{1}{\sqrt{2}}\big(\ket{+} + \ket{-}\big)~,\\[0.4cm]
\ket{y}&=&\frac{i}{\sqrt{2}}\big(-\ket{+} + \ket{-}\big)~.
\end{array}
\label{eq:XYtoPM}
\end{equation}
One photon of the pair is sent to Alice and one to Bob, as showed in Fig. \ref{fig:GE}(a). Alice chooses whether to perform polarization measurement in the circular base (setting $C$) or in the linear base (setting $L$). Bob, on the other hand, prepares a two-beam direct interferometer setup \cite{Kalbarczyk2019, Alamri2019, Peter2020, Lasagni2017}. Each photon that reaches Bob passes through a polarization beam splitter (PBS) in the circular base. Hence, states $\ket{+}$ are pushed toward the upper channel, while states $\ket{-}$ are pushed toward the lower channel. A phase shift generator (PSG) is placed in the upper channel, by which the state $\ket{+}$ acquires a phase $\alpha$, where $\alpha$ is set as needed. Next, a spin flip operator (SFO) is placed in the upper channel so to flip the circular polarization, thereby transforming states $\ket{+}$ into $\ket{-}$. A set of mirrors conveys the two photon beams into the measurement point ($F$), where a small photodetector is placed. Since the two beams are in phase, they are able to interfere at the point $F$. For a successful interference, however, the polarization changes and phase shifts induced by reflections must be taken into considerations. Finally, Alice and Bob agree that Alice's measurement on her photon happen slightly before Bob receives his photon. This can be easily fulfilled by adjusting the distances between Bob, Alice and the source $S$.

\subsection{Setting $C$}
\label{sec:SettingC}
Alice chooses to perform polarization measurements of her photons in the circular basis. From the first row in Eq. \eqref{eq:State}, and by following the Born rule for probabilities, one derives that the probability for Alice to obtain $\ket{+}$ or $\ket{-}$ on each of her photons is equal to 1/2. Alice's polarization measurements will therefore be peaked around 0 --- i.e. she is going to obtain as many photons polarized $\ket{+}$ as many polarized $\ket{-}$ ---, following a binomial distribution \cite{Fratini2011}.
As a consequence of Alice's measurements, the quantum state of the photon pair collapses: whenever Alice measures $\ket{+}$, the photon that reaches Bob acquires the state $\ket{-}$, and vice versa. Such a collapse is supposed to happen instantaneously, even in the case Bob and Alice were far away from each other. 

Although each of Bob's photons is in a pure quantum mechanical state, Bob has no knowledge on which state each photon possesses. This entails that the whole ensemble of photons that reaches Bob is a completely mixed ensemble of states $\ket{+}$ and $\ket{-}$. 

Let us follow the path of one of Bob's photons, while looking at Fig. \ref{fig:GE}. Let us suppose that, after the collapse that follows Alice's measurement, Bob's photon's state is $\ket{+}$. 
At the PBS, the photon is pushed toward the upper channel with 100\% probability. Next, it acquires a phase shift given by the PSG. The photon's polarization state is then flipped to $\ket{-}$ by the SFO. Finally, the photon is redirected to the point $F$ by the mirrors. The probability to find the photon at the point $F$, once the photon has run for the whole path described above, is certainly 100\%. Bob's photodetector $F$ will therefore `click', thereby signaling that a photon has been detected. On the other hand, in case the initial photon polarization state were $\ket{-}$, the photon path would be simpler: the photon would be pushed toward the lower channel by the PBS with 100\% probability, then be redirected to the point $F$ by the mirrors, then be detected with 100\% probability by $F$, using hypothesis \ref{hyp}.2. Since not in any case is the photon running on the upper and lower channel at the same time, there won't ever be any interference pattern at the detector location $F$, in line with Fig. \ref{fig:headon}(c) and (d).

\subsection{Setting $L$}
\label{sec:SettingL}
Alice chooses to perform polarization measurements of her photons in the linear basis. From the second row in Eq. \eqref{eq:State}, and by following the Born rule for probabilities, one derives that the probability for Alice to obtain $\ket{x}$ or $\ket{y}$ on each of her photons is equal to 1/2. Thus, similarly to before, Alice's polarization measurements will be peaked around 0 --- i.e. she is going to obtain as many photons polarized $\ket{x}$ as many polarized $\ket{y}$ ---, following a binomial distribution \cite{Fratini2011}.
As a consequence of Alice's measurements, the quantum state of the photon pair collapses: whenever Alice measures $\ket{x}$, the photon that reaches Bob acquires the state $\ket{y}$, and vice versa. We shall call these two possibilities as `case $y$' and `case $x$', respectively.

Although each of Bob's photons is in a pure quantum mechanical state, Bob has no knowledge on which state each photon possesses. The photons that reach Bob are therefore a completely mixed ensemble of states $\ket{x}$ and $\ket{y}$. 

The density matrices of setting $C$ and $L$ are the same, as can be easily calculated. Since the density matrix is supposed to contain the whole information pertaining to the state, Alice should in principle not be able to make use of the two different settings to send information to Bob. The terms of the NC theorem normally end here with such a statement. To this extent, our thought experiment complies with the NC theorem. Nevertheless, we shall try to highlight a measurable difference between settings $C$ and $L$ that Bob and Alice can use to exchange information.

\begin{figure}[t]
\includegraphics[scale=0.7]{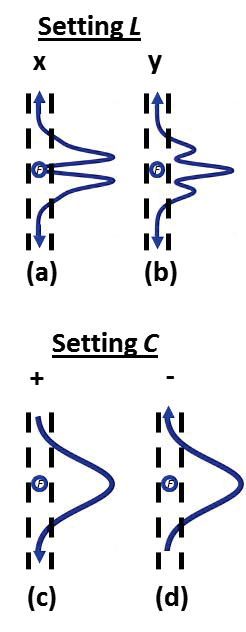}
\caption{Interference region, for $\alpha=\pi$, at time $t=t_0$, depending on Alice's setting and cases. The red-shaded region is monitored by the detector.
Setting $L$: 
	(a) The photon that reaches Bob is $\ket{x}$. 
	(b) The photon that reaches Bob is $\ket{y}$.
Setting $C$: 
	(c) The photon that reaches Bob is $\ket{+}$.
	(d) The photon that reaches Bob is $\ket{-}$.
}
\label{fig:headon}
\end{figure}

Let us follow the path of one of Bob's photons in setting $L$, while looking at Fig. \ref{fig:GE}(a). Let us suppose that, after the collapse following Alice's measurement, we have a `case $x$', i.e. Bob's photon's state is $\ket{x}$, as written in Eq. \eqref{eq:XYtoPM}. Let us denote with $t=0$ the time at which Bob's photon enters the interferometer, while with $t=1$, $2$, $3$, $4$ we denote the times when the photon exits the PBS, PSG, SFO, and is redirected to the point $F$ by the set of mirrors, respectively. At the PBS, the circular components are pushed to the different channels. At the exit of the PBS, Bob's photon state is thus
\begin{equation}
\ket{\gamma_x(t=1)} = \frac{1}{\sqrt{2}}\big(
	\ket{+} \ket{\textrm{\small upper}} 
	+ 
	\ket{-} \ket{\textrm{\small lower}}
	\big)~,
\label{eq:PBS}
\end{equation}
where $\ket{\textrm{\small upper}}$, $\ket{\textrm{\small lower}}$ denote the spacial location of the photon as upper and lower channel, respectively. The subscript ${}_x$ denotes that the state at $t=0$ happened to be $\ket{x}$. 
The phase induced by the free evolution has not been considered here nor shall be considered hereafter, since it is shared by both beams when they cross each other, at the point $F$.
After the PSG, Bob's photon state is
\begin{equation}
\ket{\gamma_x(t=2)} = \frac{1}{\sqrt{2}}\big(
	e^{i\alpha}\ket{+} \ket{\textrm{\small upper}} 
	+ 
	\ket{-} \ket{\textrm{\small lower}}
	\big)~,
\label{eq:PSG}
\end{equation}
while after the SFO the state is
\begin{equation}
\ket{\gamma_x(t=3)} = \frac{1}{\sqrt{2}}\big(
	e^{i\alpha}\ket{-} \ket{\textrm{\small upper}} 
	+ 
	\ket{-} \ket{\textrm{\small lower}}
	\big)~.
\label{eq:SFO}
\end{equation}
Finally, the set of mirrors brings the two beams together head-on at the detector region. To write the amplitude at $z=0$, one needs to sum up the amplitudes of each single beam \cite{Kalbarczyk2019, COW, Sakurai1994}. The probability to detect the photon at $z=0$ is thus
\begin{equation}
\big|\braket{\gamma_x(t=4)}{\gamma_x(t=4)}\big|^2=\frac{1}{2}\big|1+e^{i\alpha}\big|^2 = 2\cos^2\frac{\alpha}{2}~.
\label{eq:ProbX}
\end{equation}
Wavefunctions of upper and lower beam interfere. As a matter of fact, from equation \eqref{eq:ProbX} one reads that, in case the initial state of Bob's photon were $\ket{x}$, and if Bob sets $\alpha = \pi$, the probability density to find the photon at point $z=0$ would be zero, which is a consequence of destructive interference at $x=0$. Other regions of space, however, must compensate for that, in order to preserve energy and probability, at each point in time, as showed in Fig. \ref{fig:headon}(a). 

Repeating the same calculation for an initial state of the type $\ket{y}$ yields 
\begin{equation}
\big|\braket{\gamma_y(t=4)}{\gamma_y(t=4)}\big|^2 = 2\sin^2\frac{\alpha}{2}~.
\label{eq:ProbY}
\end{equation}
The probability density to find the photon at point $z=0$ is in this case equal to 200\%, which is a consequence of constructing interference at the point $z=0$. Similarly to before, other regions of space must compensate so to preserve energy and probability, at each point in time, thus giving rise qualitatively to a pattern similar to Fig. \ref{fig:headon}(b). 

As last note, we obtained destructive (constructive) interference at $z=0$ for the case $x$ ($y$) due to the choice of $\alpha=\pi$. If we chose $\alpha=0$, the situation would be reversed.

\section{Detecting Alice's message}
\label{sec:TracesALI}
Using hypothesis \ref{hyp}.1, we here suppose, here and in the following, that the trough formed by the interference pattern at $F$ is wide enough to contain the detector. This allows us to make a statement that violates the NC theorem, as follows. In case of setting $L$ and if Bob's photon were initially $\ket{x}$, the detector at $F$ would never `click', since this latter constantly lies on a trough. This means that in setting $L$ the detector $F$ may `click' only when Bob's photon state happen to initially be $\ket{y}$, in which case it lies on a crest. In the following, for simplicity and with no impact on the conclusions, we shall suppose that the crest is high enough so that Bob will detect all his photons that happen to initially be $\ket{y}$. 

In setting $C$, on the other hand, the detector $F$ would always `click', independently of whether Bob's photon is initially $\ket{+}$ or $\ket{-}$ -- viz. independently of Alice's measurement --, since the photon is in any case crossing the whole detector, either from above or from below. We are here using hypothesis \ref{hyp}.2. Bob thus experiences different outcomes depending on Alice's settings. Such a difference may be used by Bob and Alice to instantaneously exchange information, however far they might be. 

\sss

A consequence of the discussion presented above is that there is a difference between Alice's settings $C$ and $L$ that Bob can observe. Let us suppose that Bob sets his PSG with $\alpha=\pi$. Then, in setting $C$, Bob is going to detect all photons that are sent to him at point $F$. On the other hand, in setting $L$, Bob is going to detect in $F$ only those photons that happen to be initially set as $\ket{y}$, via the collapse that follows Alice's measurement. This means that, as soon as Alice decides to measure polarizations in the linear basis, Bob starts detecting less photons at the photodetector $F$. More quantitatively, if Bob is sent 100 photons per second, the detection rate at $F$ will be 100 photons per second in setting $C$, while it would drop down to approximately 50 photons per second in setting $L$, following a binomial distribution. 

Every second, by counting photons with the photodetector $F$, Bob understands whether Alice chose setting $C$ (photon count = 100) or setting $L$ (photon count $\le$ 100). Alice can therefore transmit one bit of information per second. Bob's understanding is almost always exact. Only in case Alice, by chance, in a given second, measures 100 linear polarization states $\ket{x}$ in setting $L$, would Bob then receive 100 photons with the same polarization state $\ket{y}$. In this case, Bob would count 100 photons at $F$ for the setting $L$. Bob's result in this setting for such a second would thus coincide with his result for the setting $C$, which fact would let Bob wrongly assume that Alice's setting was $C$. However, the probability for this to happen is less than $10^{-30}$, at each second, which can be further decreased by increasing the photon rate per second.

\section{Some discussion: an intuitive explanation}
\label{sec:APosteriori}
Did we defy the No-Communication theorem? If yes, how did we defy it? At each instance of time, the probability for Bob to detect the photon at $F$, not knowing Alice's measurements, can be calculated as the average between the two possible outcomes; that is between $\ket{+}$ and $\ket{-}$ in case of setting $C$, while between $\ket{x}$ and $\ket{y}$ in case of setting $L$. The NC theorem tells us that such a probability is equal in both settings. 
In Setting $C$, however, in order for the photon to entirely go from one side of $F$ to the other, it needs to cross the detector. Hence, if hypotheses \ref{hyp}.1 and \ref{hyp}.2 are correct, independently of how much is the probability for the detector to `click' at each instant of time, the detector will eventually `click'. 

\begin{figure}[t]
\includegraphics[scale=0.75]{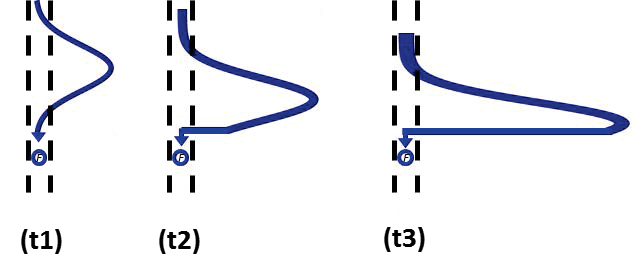}
\caption{Setting $C$, in case the photon that reaches Bob happens to be $\ket{+}$, intuitive description. Each instant of time the detector $F$ does {\it not} detect the photon implies a collapse of the photon wavefunction, by which this latter is progressively more and more packed in the region above $F$, while still be moving towards $F$ (panels (t1), (t2)). After few instants, the wavefunction will be very peaked about the point $F$ (panel (t3)), implying that the probability to find the photon at about the position $F$ is very close to 1. At that point, the detector will `click'. A similar situation occurs if the photon that reaches Bob happens to be $\ket{-}$, but in the bottom part of the channel.
}
\label{fig:Pack}
\end{figure}

To better try to understand this concept from an intuitive, non-rigorous, theoretical point of view -- i.e. why in Setting $C$ the detector $F$ would always `click' --, let us suppose Bob's photon be initially $\ket{+}$. The photon would be thus proceeding from above as in Fig. \ref{fig:Pack}(t1). For each instant of time while the detector $F$ does $not$ detect the photon, the photon wavefunction collapses, and thereby retracts in the upper part of the channel, as showed in passing from Fig. \ref{fig:Pack}(t1) to (t2) and to (t3). Once a significant portion of time is elapsed without the photon being yet detected, its wavefunction will be strongly peaked, while be still moving towards $F$, as in Fig. \ref{fig:Pack}(t3). At some point in time, the probability to detect the photon at $x\approx F$ will be close to 1: At that point the detector will `click'. In short: measurements are not instantaneous events, rather they endure in time; this allows the detector to always `click' in Setting $C$.



\section{A more rigorous treatment, the minimum size for the detector implied by quantum mechanics, and else}
Faster-than-light communication is thought be impossible, since it goes against relativity theory. This entails that one of the two hypotheses in Sec. \ref{hyp} must fall or, most probably, that both hypotheses cannot be fulfilled at the same time. In turn this implies that there must be a relation between detector size and detection efficiency implied by quantum mechanics, which we would like to explore in this section.

Suppose we have a non-relativistic massive particle, whose mass is denoted by $m$. Suppose the particle's quantum state is described by the wavefunction $\Psi(z)$ at time $t=0$. Then the quantum state evolved from time $t=0$ to $t=T$ is described by $\Psi(z,t)$ as
\begin{equation}
\Psi(z,t) = \sqrt{\frac{ m}{i\hbar 2\pi t}} \int_{-\infty }^{\infty } 
	\Psi(z')\,
	 e^{i\frac{ m (z-z')^2}{2 \hbar t}}
    \, \dd z'~.
\label{eq:evo}
\end{equation}
For $t\approx 0$, it follows $\Psi(z,t) \approx \Psi(z)$. Nevertheless, the value $\Psi(z,t)$ for $t>0$, for a certain $z$, depends on all values $\Psi(z')$, for $z'\in (-\infty, +\infty)$, where the dependence is weighted by the squared distance $(z-z')^2$, by the time that has passed $t$, and by the particle mass $m$.

We can concisely quantify this dependence in one relation as follows. Let us define $T=\sigma/v$ as the time it takes for the particle to cross a distance equal to its width $\sigma$, which is typically of the order of the wave period, and which quantifies the particle-detector interaction time. We denoted by $v$ the speed of the particle in the waveguide. Let us further denote the detector thickness by $F$ for a detector that extends within the range $z\in [-F/2,F/2]$. Then from Eq. \eqref{eq:evo}, if $\frac{ m F^2}{2 \hbar T} \lesssim 1$, values $\Psi(z>F/2,T)$ will significantly depend on the values of $\Psi(z'<-F/2)$. This means that values of the wavefunction after the detector will significantly depend on and grow from the values of the wavefunction before the detector, for the duration of the interaction. In turn this means that the particle may pass through the detector unobserved, i.e. the particle may tunnel through the detector. 

By using $T=\sigma/v$ and the DeBroglie relation $\lambda = mv/\hbar$, we may cast this relation as
\begin{equation}
\epsilon \lesssim 
 \frac{1}{2} \frac{F^2}{ \lambda\sigma}
\label{eq:detEff}
\end{equation}
where $\lambda$ is the DeBroglie wavelenght and $\epsilon$ is the detector efficiency. The above relation links detector efficiency, detector size, particle width and DeBroglie wavelength.
A similar relation should hold also for the detection of massless particles. 

For a free particle, it holds $\lambda \sim \sigma$. Thus Eq.\eqref{eq:detEff} implies that the detector size $F$ must satisfy the following relation, so to allow for 100\% efficient detector ($\epsilon \simeq 1$):
\begin{equation}
F \gtrsim \sqrt{2} \lambda~.
\label{eq:detEff_fin}
\end{equation}

\smallskip 

\section{Concluding} 
Equation \eqref{eq:detEff_fin}, or its massless counterpart, can be used to invalidate our hypotheses in Sec. \ref{hyp}. As a matter of fact, from Eq. \eqref{eq:detEff_fin} one reads that an efficient detector cannot be as small as needed. Rather, a perfect detector needs to be at least as large as the width of the particle times square root of two. In turn this implies that both hypotheses in Sec. \ref{hyp} cannot be fulfilled at the same time. As a consequence, conclusions leading to faster-than-light communications are invalidated by theory.

Having a detector larger than the width of the particle being detected seems a reasonable limitation to demand for a perfect detection and seems to be required by theory, viz. Eq. \eqref{eq:detEff_fin}. However, as already remarked in Sec. \ref{hyp}, one should remember that atoms are perfect detectors in nanometric waveguides (reflection coefficient saturates to 1), and yet are four orders of magnitude smaller than the wavelength of the photon being detected. Therefore those limitations seem not to apply to nature. As a consequence, conclusions leading to faster-than-light communications are eventually {\it not} invalidated by experiments.

Thus the question remains (at least to us) open. 

\section{To the reader} 
As long before mentioned, we do not claim to have found superluminal communication. Furthermore, papers that propose faster-than-light communication, or any possibility for it, are usually not a topic for the research enterprise. Thus this work is not meant to be published. Rather, it is meant to be bread for thoughts, for researchers who are interested in foundations of quantum mechanics.


\end{document}